\documentclass[fleqn,usenatbib]{mnras}

\usepackage{newtxtext,newtxmath}
\usepackage[normalem]{ulem}
\usepackage[percent]{overpic}
\newcommand{\MT}[1]{{\color{black}{~#1}}}

\newcommand{\lcdm}{$\Lambda$CDM}

\newcommand{\Omm}{\Omega_{\rm m}}

\newcommand{\Ads}{A_{\rm ds}}
\newcommand{\ns}{n_{\rm s}}
\usepackage[T1]{fontenc}

\DeclareRobustCommand{\VAN}[3]{#2}
\let\VANthebibliography\thebibliography
\def\thebibliography{\DeclareRobustCommand{\VAN}[3]{##3}\VANthebibliography}

\usepackage{graphicx}	
\usepackage{amsmath}	

\title[DES Y3 + BOSS Full Shape]{Interacting dark energy constraints from the full-shape analyses of BOSS DR12 and DES Year 3 measurements}

\author[M. Tsedrik et al.]{
M. Tsedrik$^{1, 2}$,\thanks{mtsedrik@ed.ac.uk} 
S. Lee$^3$,
K. Markovic$^3$, 
P. Carrilho$^1$,
A. Pourtsidou$^{1,2}$, 
C. Moretti$^{4, 5, 6}$, B. Bose$^{1,2}$,
\newauthor E. Huff$^3$, A. Robertson$^7$, P. L. Taylor$^{8,9,10}$, J. Zuntz$^1$
\\
$^1$Institute for Astronomy, University of Edinburgh,  Royal Observatory, Blackford Hill, Edinburgh, EH9 3HJ, UK\\
$^2$Higgs Centre for Theoretical Physics, School of Physics and Astronomy, Edinburgh, EH9 3FD, UK\\
$^3$Jet Propulsion Laboratory, California Institute of Technology, 4800 Oak Grove Drive, Pasadena, CA 91109, USA \\
$^4$SISSA - International School for Advanced Studies, Via Bonomea 265, 34136 Trieste, Italy \\
$^5$Centro Nazionale `High Performance Computer, Big Data and Quantum Computing'\\
$^6$INAF - Osservatorio Astronomico di Trieste, Via Tiepolo 11, I-34143 - Trieste, Italy \\
$^7$Observatories, Carnegie Institution for Science, 813 Santa Barbara Street, Pasadena, CA 91101, USA\\
$^8$Center for Cosmology and AstroParticle Physics (CCAPP),
The Ohio State University, Columbus, OH 43210, USA\\
$^9$Department of Physics, The Ohio State University, Columbus, OH 43210, USA\\
$^{10}$Department of Astronomy, The Ohio State University, Columbus, OH 43210, USA}

\date{Accepted XXX. Received YYY; in original form ZZZ}

\pubyear{\the\year{}}

\begin{document}
\label{firstpage}
\pagerange{\pageref{firstpage}--\pageref{lastpage}}
\maketitle

\begin{abstract}
Dark Scattering (DS) is an interacting dark energy model characterised by pure momentum exchange between dark energy and dark matter. It is phenomenologically interesting because it is unconstrained by CMB data and can alleviate the $S_8$ tension. We derive constraints on cosmological and DS parameters using three two-point correlation functions from the Dark Energy Survey third year data release (DES Y3). We then add information from the multipoles of the galaxy power spectrum combined with BAO measurements using the twelfth data release of the Baryon Oscillation Spectroscopic Survey (BOSS DR12) and external BAO measurements. We compare results from the direct combination of the probes with the joint posterior distribution calculated with a normalising flow approach. Additionally, we run a CMB analysis with the Planck Public Release 4 (PR4) for comparison of the cosmological constraints. Overall, we find that the combination of probes allows minimising projection effects and improves constraints without the need to include CMB information. It brings the marginalised posterior maxima closer to the corresponding best-fit values and weakens the sensitivity to the priors of the spectroscopic modelling nuisance parameters. These findings are highly relevant in light of forthcoming data of surveys like DESI, Euclid, and Rubin.
\end{abstract}




\section{Introduction}

Data from the latest generation of cosmological surveys, such as DESI\footnote{\url{https://www.desi.lbl.gov}}, Euclid\footnote{\url{https://www.euclid-ec.org}} and Rubin\footnote{\url{https://rubinobservatory.org}} will shed light on the nature of the dark sector (dark energy and dark matter) by providing Large Scale Structure (LSS) information with unprecedented precision.
These surveys will allow us to draw unambiguous conclusions about the current cosmological tensions, for example the so-called $S_8$ tension: a $\sim$$2\sigma$ disagreement in the amplitude of matter fluctuations between the one measured in LSS experiments \citep{kids2021, des2022} and the one measured with the CMB and propagated to later times using the standard cosmological model \citep{planck2018cosmo}. 

Dark Scattering \citep[DS, ][]{Simpson:2010vh, Pourtsidou:2013nha} is one of the potential `new physics' solutions for the $S_8$ tension. It is a subclass of models in which dark energy is coupled to cold dark matter via a momentum-exchange interaction. Theoretical modelling for observables in the DS model has been extensively tested on simulations and applied to data \citep{Pourtsidou:2016ico, Kumar:2017bpv, Bose:2017jjx, Bose:2018zpk, Carrilho:2021hly, Carrilho:2021otr, ManciniSpurio:2021jvx,  Linton:2021cgd, Tsedrik:2022cri, Carrilho:2022mon, Carrion:2024itc}. In this work, we provide constraints on DS cosmology with DES Y3, a photometric dataset consisting of three two-point correlation functions: cosmic shear, galaxy clustering, and the cross-correlation of source galaxy shear with lens galaxy positions. We then showcase the power of combining it with a full-shape analysis using data from BOSS DR12, a spectroscopic dataset consisting of three power spectrum multipoles and BAOs. For reference, we also provide constraints on the standard  cosmological model, \lcdm. 

\section{Model}
\label{sec:model}

Dark Scattering is an interacting dark energy model,
in which dark energy and cold dark matter exchange momentum via elastic scattering \citep{Simpson:2010vh, Pourtsidou:2013nha, Skordis:2015yra}. No energy transfer between those two species is involved in the scattering process, hence leaving the expansion history unaffected by the interaction. Additionally, this form of coupling only weakly affects the CMB and is in remarkable agreement with Planck measurements \citep{Pourtsidou:2016ico}.     
The DS model is characterised by two extra parameters with respect to $\Lambda$CDM: the equation of
state parameter, $w$, and the interaction parameter, $\xi$. The interaction parameter is defined as the ratio between the scattering cross section and the dark matter mass. For simplicity we take the dark energy equation of
state parameter to be constant.  
For the case of constant $w$, the Friedmann equation in a flat universe is given by
\begin{equation}
\label{eq:H}
    H^2 = H_0^2 \left( \Omm~ a^{-3} + \Omega_{\rm DE}~ a^{-3(1+w)} \right)  \, ,
\end{equation}
where $H = \dot{a}/a$ is the Hubble rate with the scale factor $a=1/(1+z)$, the dot denotes a derivative with respect to cosmic time, and $H_0$ is the value of the expansion rate today; $\Omm$ is the matter density parameter today and $\Omega_{\rm DE}=1-\Omm$ is the dark energy density parameter. The total matter fraction includes contributions from cold dark matter ($\Omega_{\rm c}$), baryons ($\Omega_{\rm b}$) and massive neutrinos ($\Omega_\nu$). 

The interaction between dark matter particles and the dark energy fluid manifests itself through an additional friction or drag in the movement of dark matter particles. Under the assumption of  the dark energy speed of sound being equal to the speed of light, we can write the linearised Euler equation as
\begin{equation}
\label{eq:Euler}
    a~\partial_a \Theta + \left( 2 + (1+w)~\xi~ \frac{\rho_\mathrm{DE}}{H} + \frac{a~\partial_a H}{H} \right) \Theta + \frac{\nabla^2 \Phi}{a^2 H^2} = 0 \, .
\end{equation}
Here $\partial_a$ is the derivative with respect to the scale factor, $\Theta = \theta_{\rm c} /a H$ with $\theta_\mathrm{c}$ the dark matter velocity divergence, $\Phi$ is the corresponding gravitational potential, and $\rho_{\rm DE}$ is the dark energy density. We evolve baryons and dark matter as a single species. 

From \autoref{eq:H} and \autoref{eq:Euler} we find the linear growth factor, $D$, and the growth rate $f = {\rm d} \ln D / {\rm d} \ln a$ by solving the linearised growth equation. In \autoref{eq:Euler}, we see that if $w \approx -1$ the drag/friction-term becomes negligible and the scattering parameter $\xi$ becomes unconstrained. For this reason,  
in our analysis we vary the combined parameter $\Ads=(1+w)\xi$,
with a well-defined \lcdm-limit of $\Ads=0$ b GeV$^{-1}$ and $w=-1$. Since 
$\xi$ is always positive (or zero), the sign of $\Ads$ and $(1+w)$ has to be the same. The parameter space region that can resolve the $S_8$ tension corresponds to $\Ads>0$ and $w>-1$. 
In this case, the linear growth factor decreases with respect to the \lcdm ~scenario at low redshifts. This leads to a scale-independent suppression of the power spectrum at large scales in DS. However, on smaller scales, in collapsed objects, dark matter particles experience additional friction, which leads to a scale-dependent enhancement of structure growth. Such nonlinear behaviour is modelled within the halo model reaction framework \citep{Cataneo:2018otr, Bose:2020otr}, which was tested against the DS N-body simulations presented in \citet{Baldi:2014ica,Baldi:2016zom} \citep[see also][]{Palma:2023ggq}.

\section{Data and methods}
\label{sec:data_and_methods} 

We consider the publicly released measurements of the three two-point correlation functions (3$\times$2pt) from DES Y3 \citep{DES:2021wwk, DES:2022ccp}. They contain cosmic shear, galaxy clustering and galaxy-galaxy lensing information from sources in four redshift-bins and lenses from the first four redshift-bins of the MagLim sample \citep{DES:2020ajx}. The corresponding covariance matrix is obtained analytically as described and validated in \citet{DES:2020ypx}. We use the same scale-cuts as in the DES Y3 \lcdm ~baseline analysis since we apply the halo model reaction approach, a robust prescription of nonlinearities for DS \citep{Carrilho:2021otr}. 
The study of these scale-cuts is presented in \citet{DES:2021rex} and justifies the fact that we do not model effects of baryonic feedback on small scales. We use the official DES-pipeline, \texttt{CosmoSIS}\footnote{\url{https://github.com/joezuntz/cosmosis-standard-library}}, in which we substitute the computation of linear and nonlinear matter power spectra by \texttt{CAMB} \citep{camb} with an emulator, \texttt{DS-emulator}\footnote{\url{https://github.com/karimpsi22/DS-emulators}} \citep{Carrion:2024itc}. This allows for a faster computation of the theoretical prediction.

In the following bullet-points we summarise the difference with respect to the analysis set-up from \citet{DES:2021wwk}.    
\begin{itemize}
    \item Modelling of the nonlinear power spectrum is done with \texttt{HMCode2020} \citep{Mead:2020vgs} for \lcdm
    ~motivated by a better agreement with numerical simulations \citep{Kilo-DegreeSurvey:2023gfr}. 
    \item The priors are identical to the ones summarised in Table~1 of \citet{DES:2021wwk}, except of $\{\Omega_{\rm b}, h, w\}$ (for instance, $w \in [-1.3, -0.7]$). For these parameters we impose narrower flat priors due to the limited parameter range of the training set used to construct the \texttt{DS-emulator}.  
    Additionally, the prior on the interaction parameter is $\Ads \in [-30, 30]$ b GeV$^{-1}$, motivated by the constraints found in \citet{Carrilho:2022mon, Carrion:2024itc}. 
    \item We fix the total mass of one massive and two massless neutrinos to $M_\nu=0.06$ eV. This analysis choice is motivated by the weak constraints from Stage III surveys \citep{DES:2021wwk, Moretti:2023drg}. Moreover, varying $M_\nu$ introduces additional prior volume effects \citep[][]{DES:2021bpo}. 
    \item We do not include information from the small-scale shear ratio \citep{DES:2021jzg} motivated by \citet{Arico:2023ocu}.
\end{itemize}

\begin{table*}
  \scriptsize
	\centering
	\caption{Mean values and 68\% c.l. values for DS and \lcdm ~with fixed
    one massive neutrino with $M_{\nu}=0.06~{\rm eV}$ 
    for the different probes and their combination considered in this work. We show the MAP values in parentheses, and include derived constraints on $\sigma_8$. We report the number of data-points $N_{\rm data}$, the number of parameters $N_{\rm par}$ varied in the MCMC and number of analytically marginalised parameters in parentheses, $\chi^2$ and $\chi^2_{\rm prior}$ computed with MAP-values and log-evidence $\log{\mathcal{Z}}$ from the sampler. Unconstrained parameters are denoted by a dash-line. The BBN prior is applied in the FS+BAO and joint analyses.}
	\label{tab:constraints}
	\begin{tabular}{c | cc | cc | cc |}	
		 & \multicolumn{2}{c|}{{\bf 3$\times$2pt from DES Y3}} & \multicolumn{2}{c|}{{\bf FS + BAO from BOSS DR12 and external BAOs}} & \multicolumn{2}{c|}{{\bf joint}}  \\ 
         $N_{\rm data}$ & \multicolumn{2}{c|}{462} & \multicolumn{2}{c|}{456+(8+4)} & \multicolumn{2}{c|}{930} \\ 
         \hline 
		& \lcdm & DS & \lcdm & DS & \lcdm & DS \\
		\hline
		$\Omm$ & $0.305 \pm 0.024$ (0.317) & $0.303\pm 0.029$ (0.342) & $0.311^{+0.012}_{-0.014}$ (0.306) & $0.297\pm 0.016$ (0.309 | 0.311) &$0.316\pm 0.011$ (0.311) &$0.311\pm 0.014$ (0.317) \\
        $S_8$ & $0.803 \pm 0.018$ (0.799) & $0.811^{+0.031}_{-0.039}$ (0.799) & $0.746\pm 0.048$ (0.842) & $0.639^{+0.043}_{-0.063}$ (0.711 | 0.864) &$0.792\pm 0.015$ (0.808) &$0.790\pm 0.018$ (0.814)\\
        $\sigma_8$ & $0.799^{+0.039}_{-0.047}$ (0.777) & $0.811^{+0.057}_{-0.076}$ (0.748) & $0.733\pm 0.050$ (0.833) & $0.643^{+0.043}_{-0.063}$ (0.701 | 0.849) & $0.773\pm 0.024$ (0.794) &$0.775^{+0.024}_{-0.028}$ (0.792)\\
        $w$ & -1 & $-1.02^{+0.14}_{-0.19}$ (-0.81) & -1 & $-1.17^{+0.13}_{-0.08}$ ($\approx$-1 | -0.94) &-1 & $-1.04^{+0.10}_{-0.08}$ (-0.93) \\
        $\Ads$ & 0 & $-5^{+13}_{-20}$ (6) & 0 & $-21^{+16}_{-10}$ (-22 | 0) &0 & $-0.2^{+4.6}_{-5.9}$ (0.0) \\
        \hline
        $h$& -- (0.677) & -- (0.640) & $0.681\pm 0.010$ (0.676) & $0.713^{+0.018}_{-0.024}$ (0.678 | 0.664) & $0.682\pm 0.009$ (0.680) &$0.689^{+0.016}_{-0.020}$(0.668)\\
        $\ns$ & -- (0.904) & -- (0.870) & $0.974^{+0.063}_{-0.055}$ (1.036) & $0.945\pm 0.059$ (1.022 | 1.049) &$0.957^{+0.041}_{-0.047}$ (0.977) & $0.955^{+0.041}_{-0.048}$(0.985) \\
		\hline
        $N_{\rm par}$ & 27 & 29 & 17(+32) & 19(+32)  & 39(+32) &41(+32) \\
        $\chi^2$ & 511.9 & 509.8 & 470.03  & 469.7 | 468.9 & 513.4+469.4 & 513.2+468.8\\
        $\chi^2_{\rm prior}$ & 9.5 & 9.3 & 7.99 & 7.5 | 8.3 & 19.6 & 18.9 \\
        $\log{\mathcal{Z}}$ & 5749.2 $\pm$  0.2 & 5748.8 $\pm$  0.2 &  -123.3 $\pm$   0.2 & -126.9 $\pm$ 0.2 & 5630.5 $\pm$ 0.3 & 5627.8 $\pm$ 0.2 \\
	\end{tabular}
\end{table*}

Additionally, we consider the galaxy power spectrum multipoles from BOSS DR12
\citep{alam2015, gilmarin2016, beutler2017}. 
The spectroscopic data consist of four independent sets of multipoles from two different sky cuts and two effective redshifts. 
The multipole measurements and covariance\footnote{\url{https://github.com/oliverphilcox/full_shape_likelihoods/tree/main/data}} are provided in \citet{philcox2022}. 
Complementarily, we use BAO data from BOSS DR12, 
as well as pre-reconstruction BAO measurements at low redshift from the 6DF survey \citep{beutler2011} 
and SDSS DR7 MGS \citep{ross2015}. 
We also add BAO information from high redshift
measurements 
from eBOSS DR16 \citep{desbourboux2020}.  
Further in text `FS+BAO' means the full shape (FS) analysis with the BOSS DR12 data with the inclusion of all BAOs mentioned above. 

For the BOSS DR12 FS analysis and its combination with DES Y3 we add an Effective Field Theory of LSS (EFTofLSS) \citep{baumann2012, carrasco2012,perko2016, delabella2017, DAmico:2019fhj, Nunes:2022bhn} modelling module to \texttt{CosmoSIS}. The modelling code is discussed in detail in \citet{Moretti:2023drg}. It was also extensively validated on N-body simulations in \lcdm ~\citep{oddo2020, oddo2021, rizzo2023, Tsedrik:2022cri}. The EFTofLSS approach allows us to model the power spectrum multipoles in redshift space including mildly nonlinear scales. It is based on the standard perturbation theory of structure formation and includes contributions from smaller astrophysical scales via a small number of additional parameters, so-called counterterms.

We use the same set-up of parameters and priors as in \citet{Carrilho:2022mon}. In the FS+BAO analysis the chosen priors on cosmological parameters are broad; 
the only non-flat prior we impose is a Gaussian baryon density (BBN) prior 
\citep{aver2015, cooke2018, schoneberg2019}. 
Taking into account Stage III data uncertainties, the degree of freedom in the EFTofLSS and scale-independent linear growth modification in DS (see Euler \autoref{eq:Euler}), it is sufficient to re-scale the standard perturbative components of the power spectrum multipoles by the modified growth factor and rate. We compute the growth parameters in \lcdm ~and DS scenarios using the \texttt{evogrowthpy}\footnote{\url{https://github.com/PedroCarrilho/evogrowthpy}} code.

We sample posterior distributions using \texttt{Polychord} \citep{Handley:2015fda} within \texttt{CosmoSIS} for individual LSS probes and their direct combination. In addition to directly evaluating the sum of the 3$\times$2pt and FS+BAO log-likelihoods in one MCMC run, we combine posterior distributions of individual probes together via normalising flows \citep{2019arXiv191202762P}. \MT{Normalising flows are a generative model which transforms a simple base distribution into a complex target distribution via invertible functions. This makes them a very useful tool in Bayesian inference, as they provide a flexible and tractable way to approximate complex posterior distributions \citep[see, e.g.][]{Raveri:2021wfz, Karamanis:2022ksp,Mootoovaloo:2024sao}.} We fit normalising
flows to a chain of one LSS probe, then we apply probability densities learned with normalising flows from the first chain as weights to a chain of the second probe. In the end, the re-weighted chain corresponds to the joint posterior of both probes. For the combination of posteriors via normalising flows in our analysis we use \texttt{CombineHarvesterFlow}\footnote{\url{https://github.com/pltaylor16/CombineHarvesterFlow}} \citep{Taylor:2024eqc}. An advantage of the normalising flow approach for joint analysis is its speed and efficiency after individual chains are run, and its independence from a unified likelihood implementation. However, it may fail for probes with tensions and has not been thoroughly tested on highly non-Gaussian distributions. The DS model with photometric and spectroscopic probes discussed above provides a suitable stress-test, exhibiting both challenges.  

To plot and derive constraints from the posterior distributions we use \texttt{getdist}\footnote{\url{https://github.com/cmbant/getdist}} \citep{Lewis:2019xzd}. While for the best-fit, i.e. the Maximum \textit{A Posteriori} (MAP) values, we use \texttt{iminuit}\footnote{\url{https://github.com/scikit-hep/iminuit}} \citep{iminuit}.
When the MAP value disagrees with the peak of the marginalised posterior obtained from the chain, it signifies the presence of projection effects \citep{gomezvalent2022,Hadzhiyska:2023wae}. Projection effects arise when a high-dimensional non-Gaussian posterior distribution is compressed in a lower-dimensional parameter space. Projection effects are especially prominent for extended cosmologies, in which beyond-\lcdm ~parameters are strongly degenerate with cosmological parameters \citep{Moretti:2023drg, maus2024}.

Additionally, we run a CMB analysis for DS using a modified version of \texttt{CLASS}\footnote{\url{https://github.com/PedroCarrilho/class_public/tree/IDE_DS}} and \texttt{Cobaya}\footnote{\url{https://github.com/CobayaSampler/cobaya}} with the Planck low-$\ell$ TT, the \texttt{lollipop} TE, EE and the \texttt{hillipop} TTTEEE likelihoods for the PR4 data release \citep{Tristram:2023haj}. We do not include CMB lensing information to show constraints that are as independent as possible from the late-time data from LSS.

\section{Results}
\label{sec:results}

\begin{figure*}
    \includegraphics[width=0.8\columnwidth]{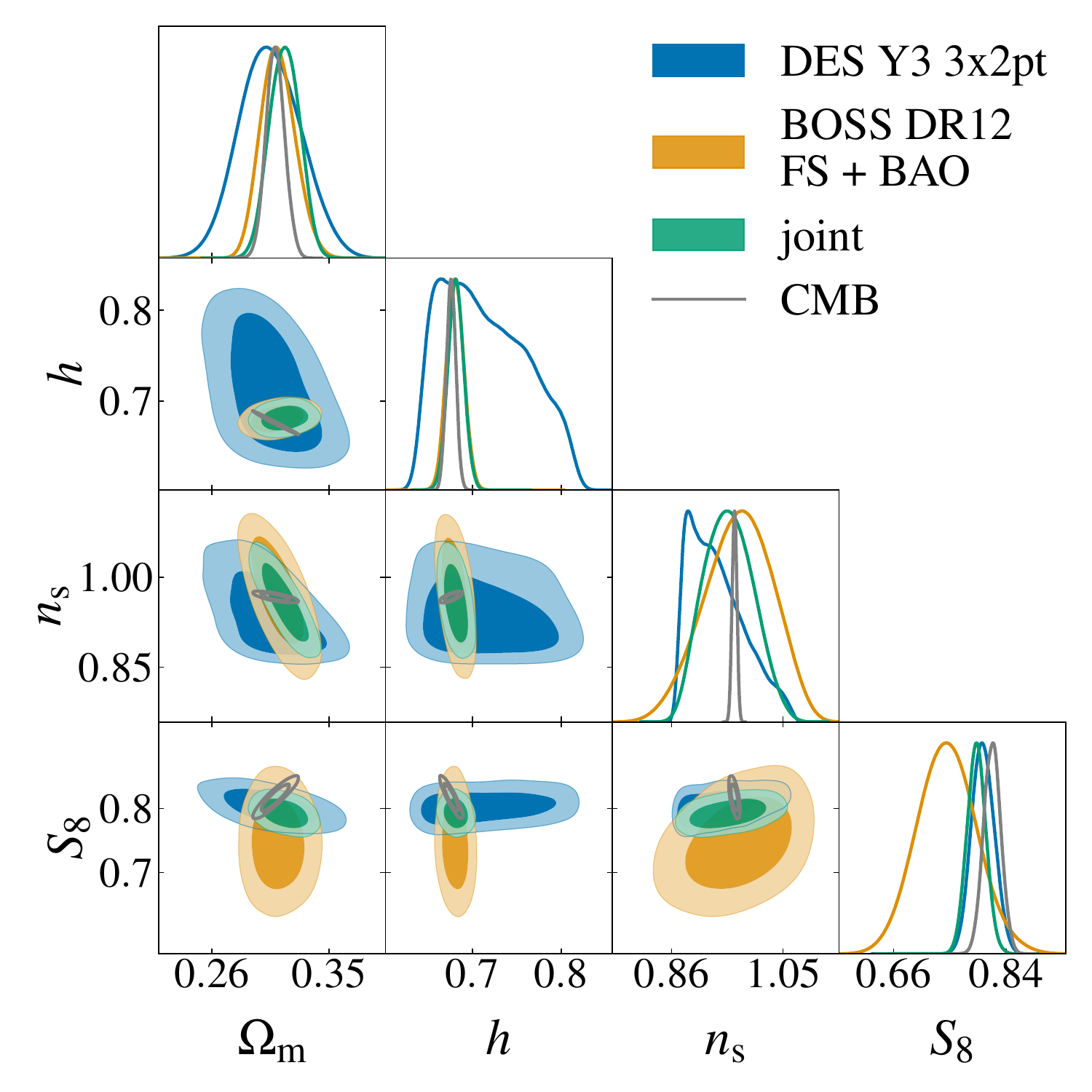}
    \includegraphics[width=0.8\columnwidth]{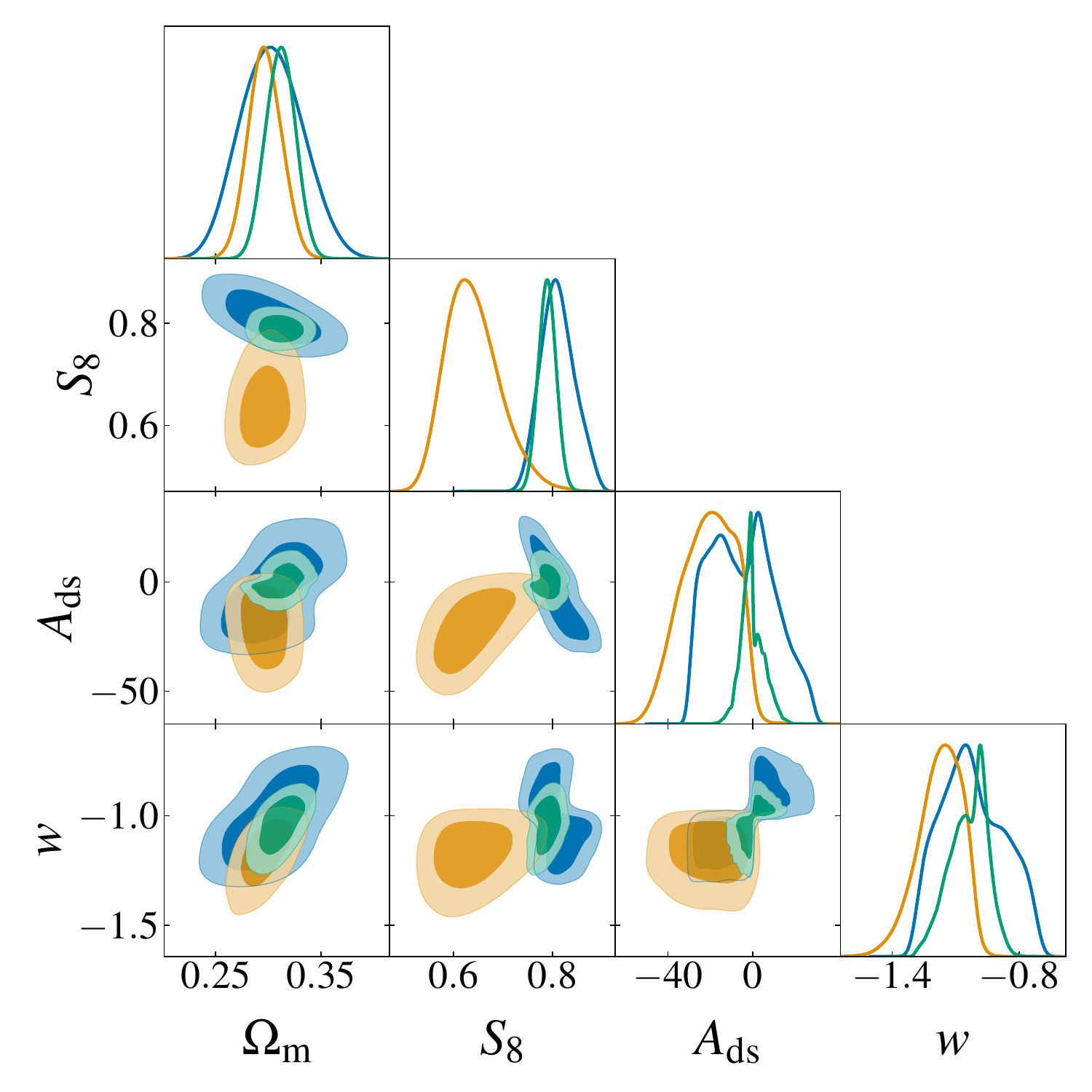}
    \caption{Marginalised posterior distribution for cosmological parameters in \lcdm ~(left panel) and DS (right panel).  Contours for DES Y3, BOSS DR12 and their combination are shown in blue, orange and green, respectively. Grey lines correspond to CMB constraints from \citet{Tristram:2023haj} without lensing information.}
    \label{fig:joint_combined}
\end{figure*}
\begin{figure}
	\includegraphics[width=\columnwidth]{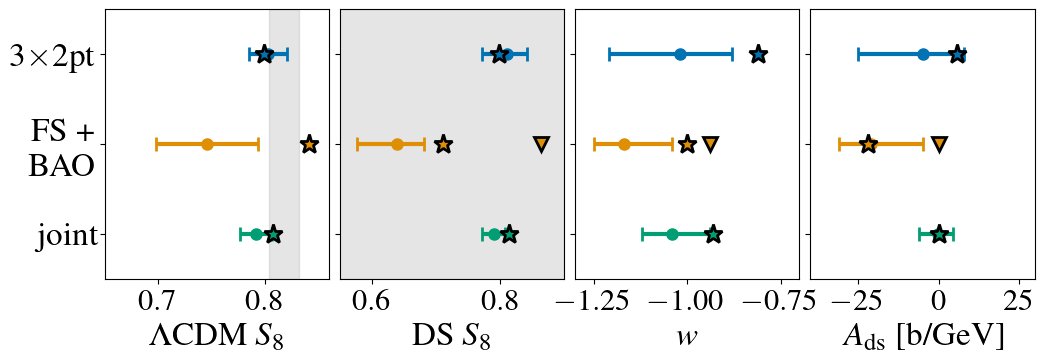}
    \caption{Projection effects in $S_8$ and the DS parameters. The error bars show the 68\% marginalised posteriors, the stars denote the MAP values, the triangles denote the second-best MAP values in FS+BAO with DS. The grey shaded line corresponds to constraints on $S_8$ from CMB data \citep{Tristram:2023haj}.}
    \label{fig:S8_tension}
\end{figure}

The constraints on cosmological parameters are given in \autoref{tab:constraints} with the means, the 68 per cent credibility intervals, and the MAP values. We do not quote the constraints on $\omega_{\rm b}$ as they are BBN-prior dominated for the FS and joint (3$\times$2pt+FS+BAO) analyses. 
 In \autoref{fig:joint_combined} we show the posterior distributions of cosmological parameters in \lcdm ~(left panel) and DS (right panel). For DS we omit $n_{\rm s}$ and $h$ because it shows a trend similar to the \lcdm ~case. Based on $\chi^2$ statistics and comparison of the Bayes factors, the performance of both models is very similar, with no clear preference for \lcdm ~or DS.

We first note that the DES-only re-analysis in \lcdm~yields  a higher value of $S_8$ than the official results in \citet{DES:2021wwk}. This is due to our analysis choices listed in \autoref{sec:data_and_methods} and in agreement with the DES shear re-analysis in \citet{Kilo-DegreeSurvey:2023gfr}. DS expands the uncertainty on $S_8$ and drives the mean value higher, while the extended parameters are prior dominated and consistent with the \lcdm-limit. However, the best-fit values of $S_8$ in both models are very close. The primordial amplitude computed at the MAP values is $\ln{(10^{10}A_s)}=2.865$ in \lcdm ~and $\ln{(10^{10}A_s)}=3.042$ in DS. The latest CMB constraint without lensing information with Planck in PR4 for \lcdm ~from \citet{Tristram:2023haj} is $\ln{(10^{10}A_s)}=3.040\pm 0.014$. Our PR4 analysis with DS constrains the primordial amplitude to $\ln{(10^{10}A_{\rm s})}=3.035\pm 0.014$. From this we see that DS offers a solution that can consistently connect early-time measurements of the matter density fluctuations in the CMB with late-time LSS measurements. Additionally, our PR4 analysis with DS constrains the following parameters: $\omega_{\rm b}=0.0223\pm 0.0001$, $\omega_{\rm c}=0.118\pm 0.001$, and $n_{\rm s}=0.968\pm 0.004$. However, it leaves the highly degenerate parameters $w, \Ads, h$ unconstrained with CMB data alone.

The FS+BAO analysis reproduces the results of ~\citet{Carrilho:2022mon} as expected with a strong shift in $S_8$ towards lower values in the marginalised posterior. Note that the authors quote best-fit values of the analytically marginalised posterior. Based on the MAP values (see \autoref{fig:S8_tension}) we see that the shift is caused by projection effects in both cosmologies; the data prefer higher values of $S_8$ than the posterior averages. Moreover, from minimising the full posterior in DS we find very close $\chi^2$ values that correspond to two very different values of the extended parameters. This implies that the data are not constraining enough to draw any conclusion about the DS constraints in this analysis setup.  

In the joint (3$\times$2pt+FS+BAO) analysis for \lcdm ~the $S_8$-value remains slightly lower than the value extrapolated from the CMB measurements 
without lensing information. For instance, the tension is quantified by $1.9\sigma$ and $1.3\sigma$ with \citet{planck2018cosmo} and PR4, respectively. The corresponding primordial amplitude computed at the MAP values equals to $\ln{(10^{10}A_s)}=2.975$. Meanwhile, DS constraints in the joint analysis are significantly improved, especially on $\Ads$. The corresponding primordial amplitude computed at the MAP values is $\ln{(10^{10}A_s)}=3.029$. Similar to the DES-only analysis, in the joint analysis the MAP values are more in agreement with the CMB in DS than in \lcdm. 

\begin{figure*}
    \includegraphics[width=0.8\columnwidth]{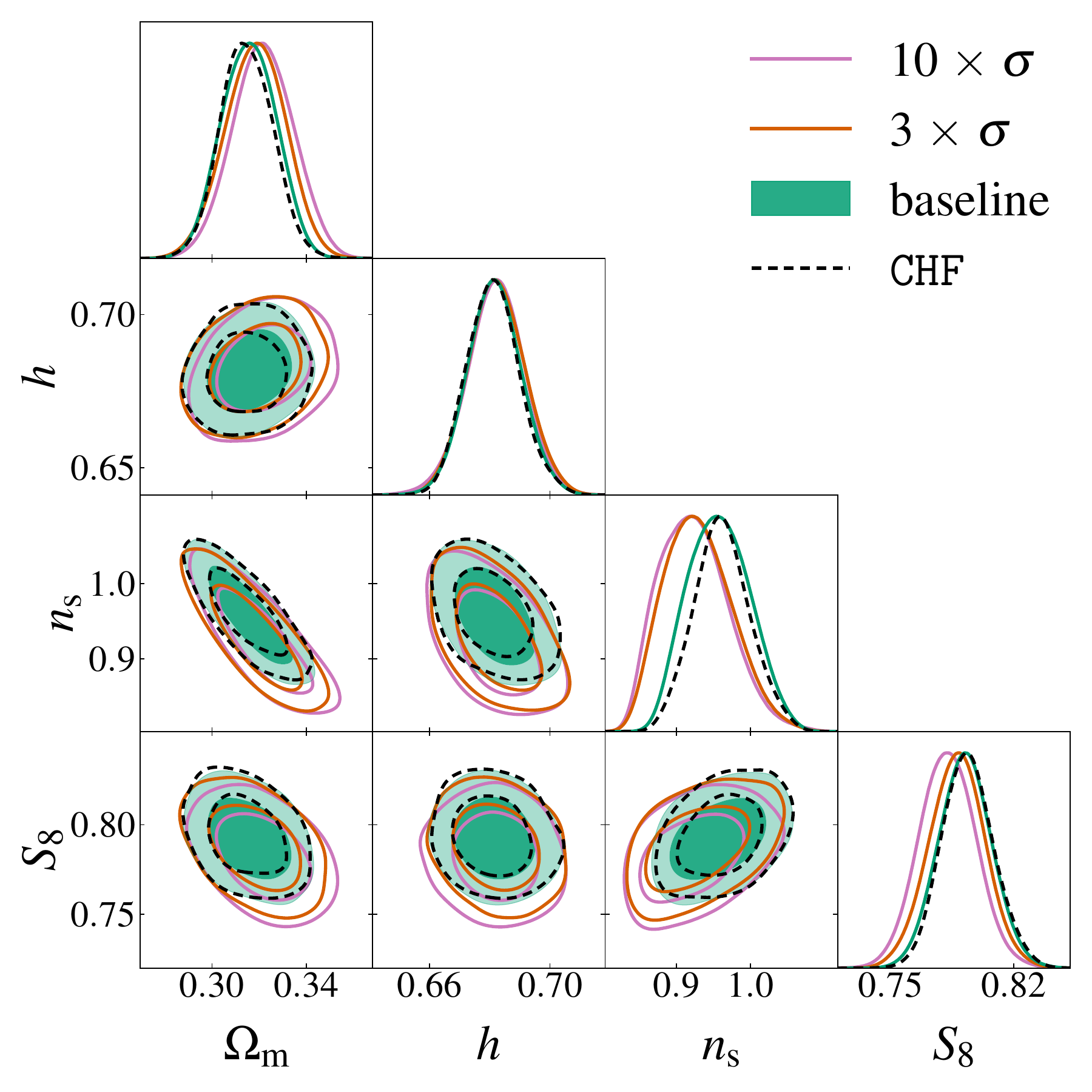}
     \begin{overpic}[scale=0.22]{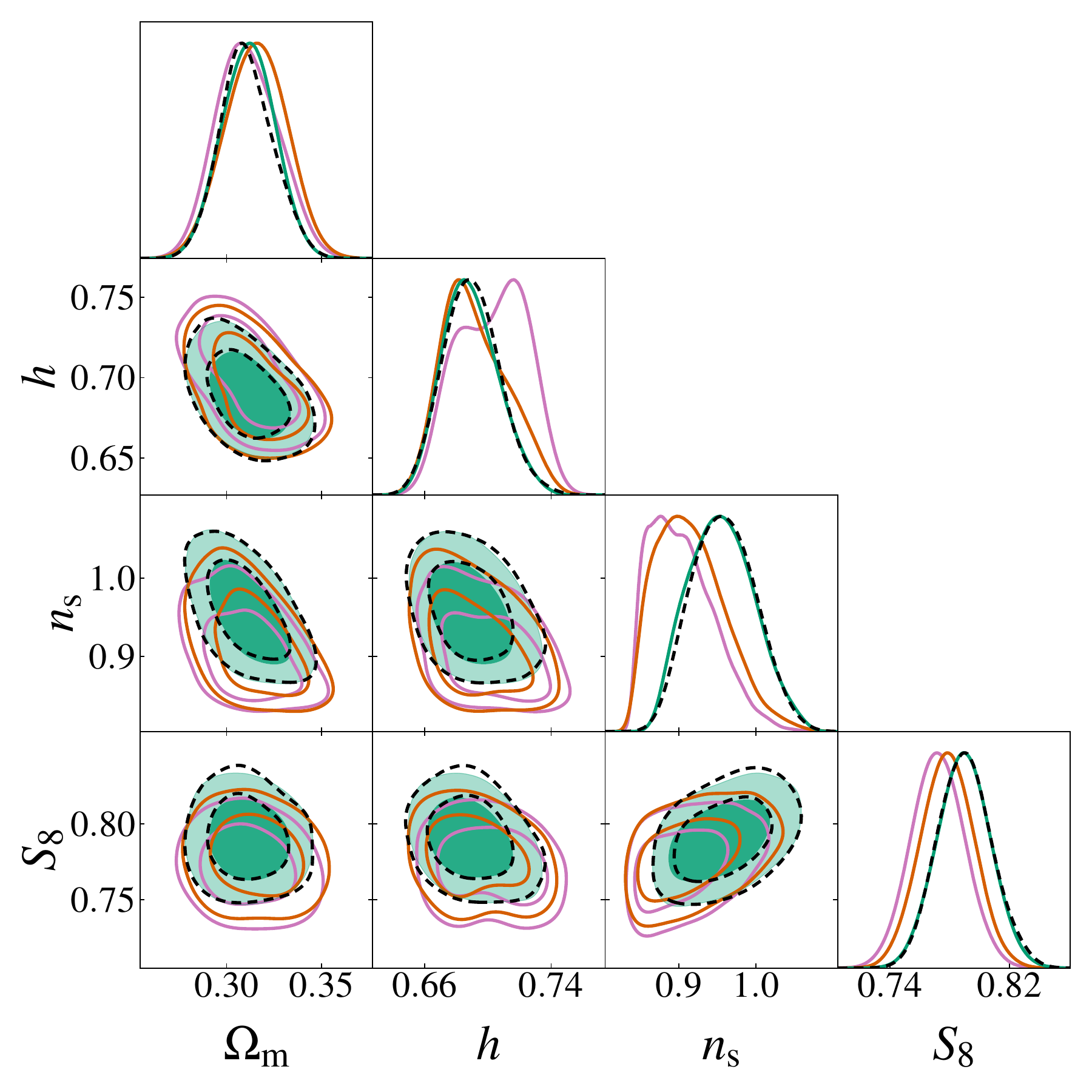}
     \put(56, 55){\includegraphics[scale=0.2]{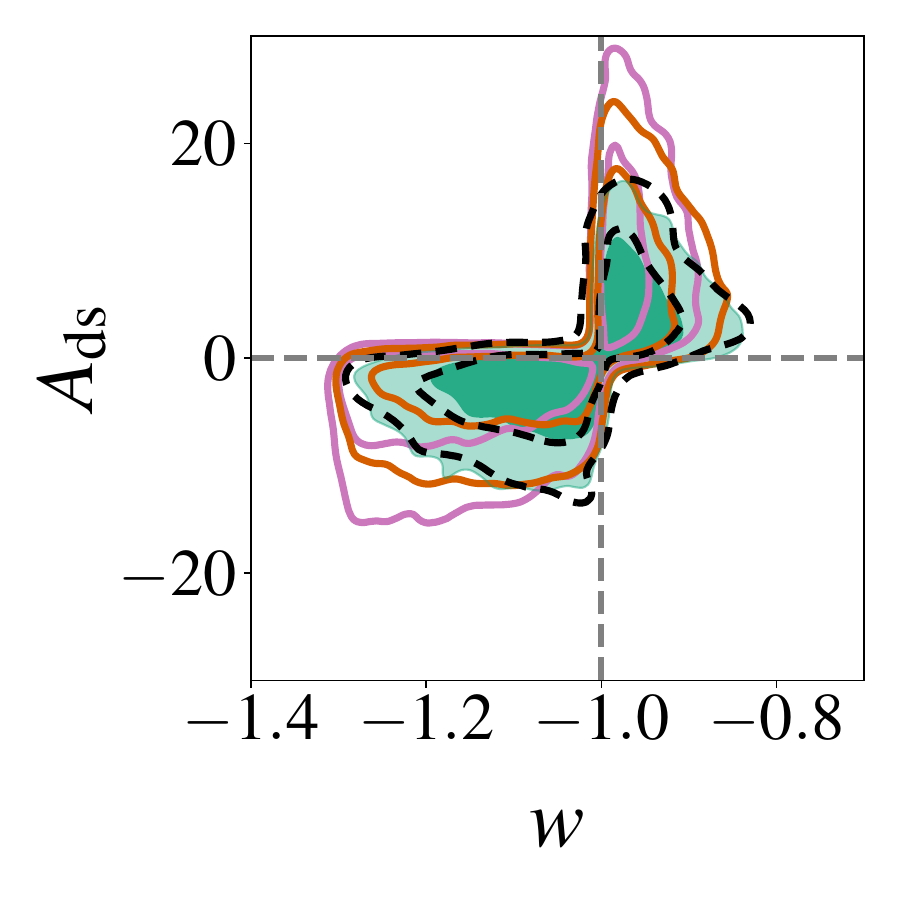}}  
  \end{overpic}
    \caption{Marginalised posterior distribution for cosmological parameters in the joint \lcdm ~(left panel) and DS (right panel) analyses for three different choices for the priors on EFTofLSS nuisance parameters: baseline case (green) and the
cases with standard deviations of nuisance parameters, except for $b_1$, increased by factors of 3 (orange) and
10 (pink). Dashed black line shows baseline joint analysis with the normalising flow, $\texttt{CombineHarvesterFlow}$ ($\texttt{CHF}$), approach.}
    \label{fig:joint_priors}
\end{figure*}

In ~\citet{Carrilho:2022mon}  the authors demonstrated that CMB-informed priors on the primordial parameters, $n_{\rm s}$ and $A_{\rm s}$, improve constraints on the extended parameters, decrease the shift in $S_8$ and the projection effects. As we show above, constraints on these parameters from the CMB in \lcdm ~and DS are similar, hence the choice of the priors is justified. We find similar trends in the combination of the spectroscopic and photometric probes without CMB-informed priors on the primordial parameters. In the joint analysis our constraints on $\Ads$ have similar uncertainty, constraints on $w$ are $\sim$3-times weaker, while $S_8$ constraints are $\sim$2-times stronger than the ones from the FS+BAO analysis with the informative priors. Additionally, in the joint analysis, the difference between the MAP and marginalised mean values is reduced (see \autoref{fig:S8_tension}), signifying a decrease in the projection effects with respect to the individual LSS probes. This is also demonstrated in \autoref{fig:joint_priors}, where we vary the standard deviation of the Gaussian prior on the EFTofLSS nuisance parameters. The effects of enlarging the prior are small in \lcdm: only lowering $n_{\rm s}$ and slightly decreasing $S_8$. In DS, the shift towards the lower prior-bound in $n_{\rm s}$ is more prominent and resembles DES-only constraints for the largest EFTofLSS priors. We also observe a slight decrease in $S_8$ and increase in $h$. However, the `butterfly'-posterior in $w-\Ads$ around its \lcdm-limit remains for all prior-choices. This implies that the photometric probes effectively constrain the amplitude of the power spectrum and help to break the degeneracy between $S_8$ and the extended parameters.  

Finally, an application of the normalising flows approach in \lcdm ~is straightforward and in agreement with the brute-force joint analysis (see \autoref{fig:joint_priors}). For DS we have two complications: (a) the overlapping region in the marginalised $S_8$-contours between the probes is small due to projection effects; (b) the `butterfly'-shaped posterior in $w-\Ads$ is highly non-Gaussian. To address these caveats we train the normalising flows on boosted posteriors: this increases the size of the training set by approximately one order of magnitude. To better avoid projection effects in $S_8$, we repeat the FS+BAO analysis with a flat prior on the primordial amplitude, $\ln{(10^{10}A_{\rm s})} \in [2.6, 4]$, motivated by the results of the DES-only analysis. We also re-run the DES analysis by imposing a BBN prior \citep[this is known to improve \texttt{CombineHarvesterFlow} performance, see ][]{DESI:2024hhd}. Overall, we reach a good agreement between the brute-force combination and the normalising flows approach with the boosted posterior distributions and adjusted $A_{\rm s}$-prior runs. \texttt{CombineHarvesterFlow} successfully passes this stress-test on highly non-Gaussian posterior distributions with strong projections.

\section{Conclusions}

  In this letter we combine photometric and spectroscopic probes from Stage III surveys: the 3$\times$2pt correlation functions of DES Y3, and the FS+BAO measurements of BOSS DR12, including external BAOs. The methods and findings of this study are important for forthcoming beyond-$\Lambda$CDM analyses by Stage IV surveys like DESI, Euclid, and Rubin, which will simultaneously provide spectroscopic and photometric datasets with unprecedented precision. 
   Currently, the significance of the apparent $S_8$ tension between the CMB and joint LSS analysis for \lcdm ~depends on the reference CMB data \citep[see, e.g.][]{Tristram:2023haj, RoyChoudhury:2024wri}. For instance, for our analysis setup the tension is reduced to $\sim$$1.3\sigma$ when compared with Planck PR4 within the standard cosmology. In case the tension would remain with Stage IV data, Dark Scattering (DS), an interacting dark energy model, presents a potential resolution. Here we provide the first DS constraints from DES Y3 data, from the joint analysis of DES Y3 and BOSS DR12, and from Planck PR4 data. The joint analysis constraints on the DS parameters are $w=-1.04^{+0.10}_{-0.08}$ and $\Ads=-0.2^{+4.6}_{-5.9}$ b GeV$^{-1}$. There is no significant detection of DS. However, in the DES-only and joint analyses, the MAP-values of the model demonstrate an exemplary solution of the tension. We also show that the joint LSS analysis without CMB information is comparable with the single-probe BOSS analysis with CMB-informed priors. The combination of probes allows minimising projection effects without the inclusion of CMB information: it brings the marginalised posterior maxima closer to the corresponding best-fit values and weakens the sensitivity to the priors of the EFTofLSS nuisance parameters. Finally, we find good agreement between the direct joint analysis and the combination of posteriors via normalising flows.

\section*{Acknowledgements}

The authors are grateful to the BOSS and DES collaborations for making their data publicly available. MT's research is supported by grant ST/Y000986/1. SL and KM are grateful for funding from JPL’s R\&TD 7x `Future of Dark Sector Cosmology: Systematic Effects and Joint Analysis' (01STRS/R.23.312.005). AP is a UK Research and Innovation Future Leaders Fellow (grant MR/X005399/1). PC’s research is supported by grant RF/ERE/221061. CM's work is supported by the Fondazione ICSC, Spoke 3 Astrophysics and Cosmos Observations (Project ID CN\_00000013). BB is supported by a UK Research and Innovation Stephen Hawking Fellowship (EP/W005654/2). PLT is supported in part by `NASA ROSES 21-ATP21-0050'. A part of this research was carried out at the Jet Propulsion Laboratory, California Institute of Technology, under a contract with the National Aeronautics and Space Administration (80NM0018D0004). \MT{For the purpose of open access, the author has applied a Creative Commons Attribution (CC BY) license to any Author Accepted Manuscript version arising from this submission.}

\section*{Data Availability}

Links and references with sources of the data and analysis pipelines are provided in the main text. 



\bibliographystyle{mnras}
\bibliography{refs} 






\bsp	
\label{lastpage}
\end{document}